\documentclass[twocolumn,showpacs,preprintnumbers,floatfix]{revtex4}
\usepackage{amssymb}
\usepackage{graphicx}
\usepackage{epsfig}
\usepackage{dcolumn}
\usepackage{bm}

\newcommand{\ba}{\begin{eqnarray}}
\newcommand{\ea}{\end{eqnarray}}
\newcommand{\ban}{\begin{eqnarray*}}
\newcommand{\ean}{\end{eqnarray*}}
\begin{document}

\title{Boson-Fermion pairing in Bose-Fermi mixtures on 1D optical lattices}
\author{X. Barillier-Pertuisel$^{1,2}$ S. Pittel$^3$ L.Pollet$^4$ P. Schuck$^{1,2,5}$ \\
{\it $^1$Institut de Physique Nucl\'{e}aire, IN2P3-CNRS, UMR8608, Orsay,F-91406, France\\
$^2$Universit\'{e} Paris-Sud,F-91406 Orsay, France \\
$^3$Bartol Research Institute and Department of Physics and Astronomy, University of Delaware, Newark, Delaware 19716, USA\\
$^4$Institut Theoretische Physik, ETH Z\"urich, CH-8093 Z\"urich, Switzerland \\
$^5$Laboratoire de Physique et Mod\'elisation des Milieux Condens\'es, CNRS
 Universit\'e Joseph Fourier, Maison des Magist\`eres, B.P. 166, 38042
Grenoble Cedex 9, France}}
\date{\today}

\begin{abstract}

Boson-fermion pairing  is considered in a discrete environment of
bosons and fully spin-polarized fermions, coupled via an
attractive Bose-Fermi Hubbard Hamiltonian in one dimension. The
results of the T-matrix approximation for particles of equal mass
and at double half filling are compared with the results of exact
diagonalization and with Quantum Monte Carlo results. Satisfactory
agreement for most quantities is found. The appearance of a
stable, weak-coupling pairing mode is also confirmed.
\end{abstract}

\pacs{03.75.Fi, 05.30.Fk} \maketitle

One of the most intriguing aspects in the field of cold atoms
involves the study of mixed Bose-Fermi (B-F) systems. Several
boson-fermion mixtures have been realized, both
in~\cite{Ospelkaus06} and without~\cite{Truscott01} an optical
lattice, and their properties have been studied. Several very
interesting phenomena unique to mixed BF systems have been
predicted, including for example the possibility of forming
composite fermions through the pairing of a boson and a fermion
\cite{KS,LSBF}. On the theoretical side, B-F mixtures have been
studied in mean-field approximation \cite{albus, fehrmann} and
with various methods for treating B-F mixtures in one dimension
(1D). This includes exact solution using the Bethe ansatz
\cite{imambekov}, bosonization techniques~\cite{Mathey} and a
Quantum Monte Carlo treatment for a B-F mixture on an optical
lattice \cite{Pollet}. There remains, however, the need for
theoretical approaches capable of reliably describing B-F mixtures
in more than one dimension~\cite{Buechler03}.

A possible theoretical approach to mixed boson-fermion systems was
proposed recently in the context of nuclear physics, to provide a
framework for describing the transition from a fermi gas (of
quarks) to one of composite fermions (nucleons), i.e. bound
three-quark states. This problem was simplified by assuming that
two of the quarks are strongly bound and form a boson. In this
way, the extremely complex in-medium three-body problem
\cite{rapp} was replaced by the much simpler two-body problem of
fermion-boson scattering, for which a T matrix approach was
developed~\cite{Storo}. An interesting result of that study was
that, due to the presence of a Fermi surface, a stable B-F branch
was created for an arbitrarily small B-F attraction. The
underlying mechanism turned out to be analogous to the formation
of stable Cooper pairs in a pure fermi-gas, though in the latter
case the pairs are boson-like, whereas here they are fermion-like.
Several interesting questions follow naturally. On the one hand,
we would like to know how reliable the information provided by the
T-matrix approach developed in \cite{Storo} is when dealing with
complex systems involving bosons and fermions. If it is found to
be acceptably reliable, we might then hope to further develop the
method for application to the variety of systems in which boson
and fermions degrees of freedom coexist, such as those that arise
in cold atomic gases.

As the next step in this program, we report in this work a study
of B-F pairing in 1D optical lattices using the above
T-matrix approach. Such systems can be treated stastistically exactly for a
fairly large number of bosons and fermions using Quantum Monte
Carlo methods \cite{Pollet}, thereby providing an appropriate
testing ground for our method.  As we will see, even though 1D is
a quite unfavorable case for the applicability of a T-matrix
approach in ladder approximation, most quantities are nevertheless
reproduced reasonably well with this approach over a large range of
coupling strengths.

To be more specific, we consider the bosons and fermions on a 1D
lattice governed by a Hubbard model hamiltonian. The B-F
interaction is assumed to be attractive and the B-B interaction to
be repulsive. We assume further that there are no interaction
among fermions, although a repulsive F-F interaction should not
alter qualitatively our conclusions. The Hubbard model hamiltonian
for such a system is given by \vspace{-0.3cm} \ba
\label{HubbardCoordSpace}
\mathcal{H}&=& -t^b \sum_{<ij>}^L b_{i}^{+}b_{j} -t^f \sum _{<ij>}^L c_{i}^{+}c_{j} \nonumber\\
&+& \frac{U_{bb}}{2}  \sum_{i}^L  b_{i}^{+} b_{i}^{+} b_{i} b_{i}
 + U_{bf} \sum_{i}^L  b_{i}^{+} c_{i}^{+} c_{i} b_{i} ~,
\ea \noindent where $b_i$ and $c_i$ are the bosonic and fermionic
annihiliation operators on site $i$, respectively. The intersite
spacing $a$ is set to unity, i.e. $a=1$. In this work we
carry out our analysis at zero temperature, i.e. $T=0$.
For the QMC simulations, we use a small but finite inverse 
temperature $\beta = 2L$ which is sufficiently close to the  
ground state. Extrapolations to larger $\beta$ do not lead 
to significantly different results.\\
\indent
In the standard T-matrix approach \cite{FW} to this problem, we
must solve the Bethe-Salpeter equation for the B-F propagator,
that is \cite{Storo}
\ba
\label{fond}
G_{\mathbf{k},\mathbf{k}^{\prime }}(\mathbf{K},E) = G_{\mathbf{k}}^{0} \left( \mathbf{K},E\right)
\delta _{\mathbf{k},\mathbf{k}^{\prime }}
\hspace{2cm} \nonumber\\ + \sum _{\mathbf{k}_1} G_{\mathbf{k}}^{0} \left(
\mathbf{K},E\right) V_{\mathbf{k},\mathbf{k_1} }^{bf}
G_{\mathbf{k}_{1},\mathbf{k}^{\prime }}(\mathbf{K},E) ~, \ea
\noindent with $G_{\mathbf{k},\mathbf{k}^\prime}(\mathbf{K},E)$
being the Fourier transform of the B-F propagator
\ban
\label{gt}
G_{\mathbf{k},\mathbf{k}^{\prime }}^{t-t^{\prime }}(\mathbf{K})=-i\theta
(t-t^{\prime })\langle \left\{ \left( b_{\frac{\mathbf{K}}{2}-\mathbf{k}}c_{%
\frac{\mathbf{K}}{2}+\mathbf{k}}\right) ^{t},\left( c_{\frac{\mathbf{K}}{2}+%
\mathbf{k^{\prime }}}^{+}b_{\frac{\mathbf{K}}{2}-\mathbf{k^{\prime }}%
}^{+}\right) ^{t^{\prime }}\right\} \rangle ~.\ean \noindent Here
$\left\{ ,\right\}$ denotes an anticommutator, $\mathbf{K}$ is the
center-of-mass  momentum of the pair and
$\mathbf{k},\mathbf{k}^\prime$ are relative momenta. The
interactions in momentum space are ($\nu = (bf,bb)$)
\vspace{-0.2cm} \ba V_{\mathbf{k},\mathbf{k}^\prime }^{\nu } =
\frac{U_{\nu }}{L} ~\delta _{\mathbf{k},\mathbf{k}^\prime} =
g_{\nu } ~\delta _{\mathbf{k},\mathbf{k}^\prime} \ea and
$G_{\mathbf{k}}^{0} \left( \mathbf{K},E\right)$ is the B-F
propagator in the HF approximation, i.e.\vspace{-0.3cm} \ba
\label{BFpropafree} G_{\mathbf{k}}^{0} \left( \mathbf{K},E\right)
= \frac{\theta (\varepsilon _\mathbf{k} ^f - \varepsilon _F ) +
\delta _{\mathbf{K}, \mathbf{k} }n_0} {E - \tilde{\varepsilon
}_\mathbf{k} ^f - \tilde{\varepsilon }_\mathbf{K-k}^b + i\eta}~,
\ea \vspace{-0.3cm} \noindent with \ban\tilde{\varepsilon
}_\mathbf{q} ^b &=& \varepsilon _\mathbf{q} ^b + \sum _{k}
V^{bf}_{\mathbf{kq},\mathbf{kq}} \Theta (\varepsilon _F
-\varepsilon _\mathbf{k}^f) +
2n_0~V^{bb}_{\mathbf{q}0,\mathbf{q}0} ~, \ean \vspace{-0.6cm} \ban
\hspace{-2.9cm}\tilde{\varepsilon }_\mathbf{k} ^f &=& \varepsilon
_\mathbf{k} ^f +V^{bf}_{\mathbf{k}0,\mathbf{k}0} n_0 ~.\ean
\noindent In these expressions, $n_0$ represents the boson
condensate, $\mathbf{p}$ ($\mathbf{h}$) refers to a fermion
momentum above (below) the Fermi surface, $\mathbf{q}$ is a boson
momentum and $\varepsilon _\mathbf{k}^{(f,b)} =  -2 t^{(f,b)} \cos
(\mathbf{k})$.
The single particle energies in 
the denominator of (\ref{BFpropafree}) are the 
fermion and boson HF energies, which are taken together with the
unrenormalized interaction, as is usual in the HF-RPA self-consistent
scheme \cite{cross1}. Screening terms which would renormalize on the 
same footing the interaction and the single 
particle self energies are not considered in this prototype study but may be included in later work.\\
Equation (\ref{fond}) can be solved analytically and the propagator
summed over relative momenta is given by \vspace{-0.1cm} \ba
\label{GKE} G(\mathbf{K},E) = \frac{G^{0}(\mathbf{K},E)}{1-g_{bf}
G^{0}(\mathbf{K},E)} ~,\ea \noindent with $G(\mathbf{K},E) = \sum
_{\mathbf{k},\mathbf{k}^\prime} G_{\mathbf{k},\mathbf{k}^{\prime
}} (\mathbf{K},E)$. For a discrete number of sites, it is also
useful to consider the equivalent diagonalization problem which
can be thought of as a B-F Random Phase Approximation in the
particle-particle channel in complete analogy with the pure
fermion case \cite{cross1}. The relevant RPA eigenvalue equation
is \ba \label{matRPA} \left(
\begin{array}{cc}
\mathcal{A}_{\mathbf{p}^{\prime }\mathbf{q}^{\prime },\mathbf{pq}}& \mathcal{B}_{\mathbf{p}^{\prime }\mathbf{q}^{\prime }} \\
\mathcal{B}_{\mathbf{h}^{\prime}0,\mathbf{pq},\mathbf{h}0} & \mathcal{C}_{\mathbf{h}^{\prime }0,\mathbf{h}0 } %
\end{array}
\right)
\left(
\begin{array}{cc}
X_{\mathbf{pq}}^{\alpha }& Y_{\mathbf{pq}}^{\rho }\\
Y_{\mathbf{h}}^{\alpha }& X_{\mathbf{h}}^{\rho }%
\end{array}
\right) = \hspace{0.5cm}\nonumber\\ \left(\begin{array}{cc}
X_{\mathbf{p}^{\prime }\mathbf{q}^{\prime }}^{\alpha }& Y_{\mathbf{p}^{\prime }\mathbf{q}^{\prime }}^{\rho }\\
Y_{\mathbf{h}^{\prime }}^{\alpha}& X_{\mathbf{h}^{\prime }}^{\rho}%
\end{array}
\right)
\left(
\begin{array}{cc}
E_\alpha & 0\\
0 & -E_\rho
\end{array}
\right) ~,\ea \ban
\mathcal{A}_{\mathbf{p}^{\prime }\mathbf{q}^{\prime },\mathbf{pq}} &=& \delta _{\mathbf{%
pp}^{\prime }}\delta _{\mathbf{qq}^{\prime }}\left( \tilde{\varepsilon} _{\mathbf{p}%
}^{f}+\tilde{\varepsilon} _{\mathbf{q}}^{b}\right)
+ a(\mathbf{q}) ~V^{bf}_{\mathbf{p}^{\prime }%
\mathbf{q}^{\prime },\mathbf{pq}}~a(\mathbf{q}^{\prime })~,\\
\mathcal{B}_{\mathbf{h}^{\prime }0,\mathbf{pq}} &=&V^{bf}_{\mathbf{h}^{\prime }0,\mathbf{pq}%
}\sqrt{n_{0}} ~a(\mathbf{q})~,\\
\mathcal{B}_{\mathbf{p}^{\prime } \mathbf{q}^{\prime},\mathbf{h}0}
&=&V^{bf}_{\mathbf{p}^{\prime }\mathbf{q}^{\prime },0\mathbf{h}} \sqrt{n_{0}}~a(\mathbf{q}^\prime )~,\\
\mathcal{C}_{\mathbf{h}^{\prime }0,\mathbf{h}0} &=& \delta
_{\mathbf{hh}^{\prime }}\left( \tilde{\varepsilon}
_{\mathbf{h}}^{f}+\tilde{\varepsilon} _{0}^{b}\right) + \delta
_{\mathbf{hh}^{\prime}} V^{bf}_{\mathbf{h}0,\mathbf{h}0}n_{0} \displaystyle ~,\ean \noindent where we
define $a(\mathbf{q})=\sqrt{1+\delta _{\mathbf{q},0}n_{0}}~$ and 
$V^{\nu}_{\mathbf{pq},\mathbf{p}^{\prime }\mathbf{q}^{\prime}}=g_\nu~
\delta _{\mathbf{p}+\mathbf{q},\mathbf{p}^{\prime }+\mathbf{q}^{\prime}}$. In
terms of the amplitudes $X$,$Y$ the propagator ($\ref{gt}$) can be
written in the following spectral representation: \vspace{-0.1cm}
\ba G_{\mathbf{k},\mathbf{k}^\prime}(\mathbf{K},E) = \sum _\alpha
\frac{\chi ^{\alpha }_\mathbf{k}\chi ^{\alpha
+}_\mathbf{k^\prime}}{E-E_ \alpha} + \sum _\rho \frac{\chi
^{\rho}_\mathbf{k} \chi ^{\rho +}_\mathbf{k^\prime}}{E+E_ \rho} ~,
\ea \noindent with $\chi ^{\alpha +}_\mathbf{k}=(Y^{\alpha
}_{\mathbf{k}}~X^{\alpha }_{\mathbf{k},\mathbf{K-k}})$ and $\chi
^{\rho +}_\mathbf{k}=(Y^{\rho }_{\mathbf{k},\mathbf{K-k}}~X^{\rho
}_{\mathbf{k}})$.
\\For a system with $N$ particles,
the propagator has poles at $E_{\rho,\alpha}=\pm (E_0^N - E_{\rho,\alpha}^{N\pm2})$,
i.e. essentially at the excitation energies of the $N\pm2$ systems. The RPA amplitudes obey the following normalization conditions:
\ba
&&Y_{\mathbf{K}}^{\alpha}Y_{\mathbf{K}}^{\alpha^{\prime}}\Theta
(\varepsilon _F - \varepsilon ^f_\mathbf{K})+ \sum _\mathbf{p}
X_{\mathbf{p,K-p}}^{\alpha} X_{\mathbf{p,K-p}}^{\alpha^{\prime}}=
\delta _{\alpha \alpha^{\prime}} ~,\nonumber\\
&&X_{\mathbf{K}}^{\rho}X_{\mathbf{K}}^{\rho^{\prime}}\Theta
(\varepsilon _F - \varepsilon ^f_\mathbf{K})+ \sum _\mathbf{p}
Y_{\mathbf{p,K-p}}^{\rho}Y_{\mathbf{p,K-p}}^{\rho^{\prime}}=
\delta _{\rho \rho^{\prime}} ~.\ea

\indent To assess the quality of our T-matrix approximation for
the in-medium B-F problem, we consider finite numbers of sites,
where exact solution is possible. We first restrict to
half-filling and to $N_B=N_F$ and $t_b=t_f$. We exactly diagonalize the 6-site
problem, i.e. $N_B=N_F=3$. The maximum size of the matrix is 318
$\times$ 318, whereas the RPA matrix has dimension 4$\times$4. The
range of $U_{bf}$ values is arbitrary. However, since we are
interested in B-F pairing, the range of $U_{bb}$ values is, in
principle, restricted to $U_{bb}\geq |U_{bf}|/2$. Smaller values
of $U_{bb}$ lead to phase separation, i.e. the B-F pairs cluster
together and occupy only half of the available space
\cite{Pollet}. On the other hand, our T-matrix approximation does
not allow $U_{bb}$-values greater than $U_{bb}\simeq |U_{bf}|$,
since $U_{bb}$ is only purely treated in our theory, i.e. only in
HF approximation. Improving on this point is possible but left for
the future. We therefore limit ourselves to $|U_{bf}|\geq U_{bb}
\geq |U_{bf}|/2$. As an intermediate value we
take $U_{bb}=\frac{3}{4}|U_{bf}|$ throughout.\\
\begin{figure}[tph]
\epsfig{file=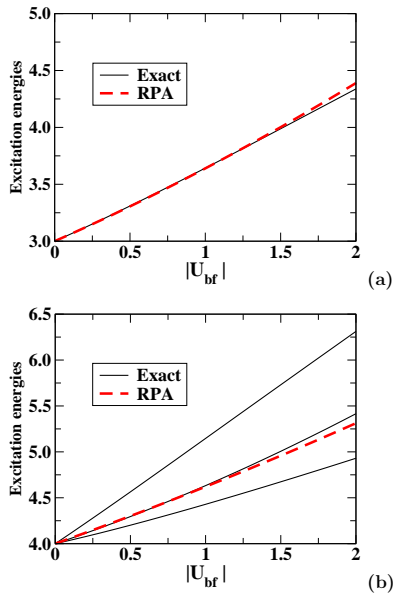,width=0.3\textwidth}
\caption{\label{excitation}(Color online) Excitation energy as a function of
$|U_{bf}|$ for the 6-site B-F Hubbard model and $\mathbf{K}=\pi/3$
(a) and for $\mathbf{K}=0$ (b). The solid lines correspond to the
results of exact diagonalization and the dashed lines to the RPA
results.}
\end{figure}

In Fig.~\ref{excitation}, we present some typical examples of
excited states. We see that the agreement between approximate and
exact excitation energies is quite satisfactory. We should note
that we have chosen examples where in the exact case there are
only low degeneracies at $U_{bf}=0$, since RPA, because of its
very low dimension, can not well reproduce a high degeneracy of
uncorrelated configurations, even if there are also cases in RPA
where in the uncorrelated limit degeneracies occur. 
However these degeneracies are always less numerous than in the exact case.
This is natural because of the dramatically reduced size of the RPA 
matrices with respect to the size of the exact ones. It should nevertheless be noted that
the RPA excitation energies somehow represents
the average trend of the bunch of exact levels.

\begin{figure}[tph]
\epsfig{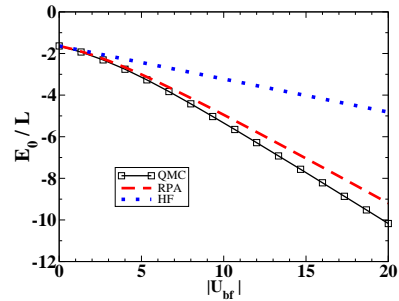}
\caption{\label{GSE70}(Color online) Ground state energy per site as a function
of $|U_{bf}|$ for $L=70$. The solid line corresponds to the QMC,
dotted line to the HF approximation and dashed line to the RPA.} 
\end{figure}

\begin{figure}[tph] 
\epsfig{file=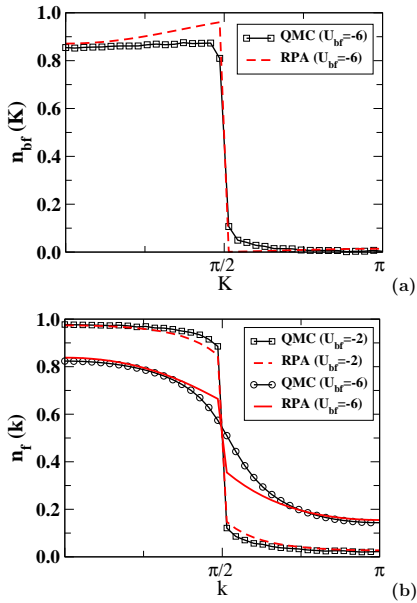,width=0.32\textwidth}
\caption{\label{number}(Color online) (a)BF occupation numbers as a
function of the total momentum $\mathbf{K}$, and (b)Fermion occupation numbers as a
function of the relative momentum $\mathbf{k}$
($U_{bb}=0.75|U_{bf}|$, $L=70$).}
\end{figure}

So far we have not invested effort to exactly diagonalize problems
with higher numbers of sites, since the dimensions of the matrices
grow exponentially. However, exact QMC results for ground state
properties for higher number of sites are available \cite{Pollet}.
For $L=70$ we show a comparison of the exact ground state energy
$E_0$ with the RPA in Fig.~\ref{GSE70} (see \cite{Storo} for the
expression for $E_0$). The error bars for QMC are smaller than the point size. 
Again the agreement is reasonable up to
rather high values of $|U_{bf}|$.
\\ \indent 
We also calculated the occupation numbers $n_{bf}(\mathbf{K}) = \sum_{\mathbf{k}\mathbf{k}^\prime} <c_\mathbf{K-k}^+ b_\mathbf{k}^+
b_\mathbf{k^\prime} c_\mathbf{K-k^\prime}>$ of the B-F pairs and compare them with the exact results 
in Fig.~\ref{number}a. In our case the B-F occupation
numbers are simply obtained from the residue of the B-F Green's function 
($\ref{GKE}$) at the poles $E_\rho$. We see that the agreement with the 
exact result is quite satisfactory. One might wonder about the upward tendency around $\mathbf{K} = \pi / 2$ in the BF 
occupation numbers for the RPA data, since it is much more pronounced than in 
the QMC data. Notice, however, that for other system parameters the 
upward trend in the occupation numbers can also be quite pronounced even 
in the exact solution. Probably, the RPA gives relatively more weight to the density 
wave correlator than to the superfluid properties because of the truncated 
model space and the fact that in three dimensions a gapped density wave was 
found in DMFT~\cite{DMFT-Titvinidze}.\\
The fermion occupation numbers $n_f(\mathbf{k}) = <c_\mathbf{k}^+c_\mathbf{k}>$ can only be obtained in a somewhat indirect
way within our formalism. This goes, however, completely parallel to what is known from the usual RPA formalism for fermions \cite{Nicole}. The
corresponding expression is given by
\ba 
\label{numb}
%n_f(\mathbf{h})&=& \sum _\rho |X_\mathbf{h}^\rho |^2 \hspace{0.5cm}\mathbf{h}\leq \mathbf{k_f} \nonumber\\
%n_f(\mathbf{p})&=& \sum _{\mathbf{K},\rho} |Y_\mathbf{p,K-p}^\rho |^2 \hspace{0.5cm}\mathbf{p}> \mathbf{k_f}
n_f(\mathbf{k}) = \sum _\rho |X_\mathbf{k}^\rho |^2 \Theta
(\varepsilon _F - \varepsilon ^f_\mathbf{k}) + \sum
_{\mathbf{K},\rho} |Y_\mathbf{k,K-k}^\rho |^2 \Theta (\varepsilon
^f_\mathbf{k} - \varepsilon _F) \ea \noindent One can demonstrate 
that ($\ref{numb}$) conserves fermion particle number. In
Fig.~\ref{number}b the fermion occupation numbers are shown for
two cases of the coupling $U_{bf}=-2,-6$. Again the agreement with
the exact case is rather good, in spite of the fact that for
$U_{bf}=-6$ our solution shows a small Fermi step whereas the
exact one seems to be completely smooth. One should realize,
however, that our theory can not describe a non-Fermi liquid
behavior and this seems to be the case for stronger interactions
in the exact solution. This deficiency of our approach to describe
specific 1D features shows up more dramatically for the boson
occupation numbers where we still obtain a large fraction of
particles condensed into the $\mathbf{q}=0$ state whereas the
exact solution is, of course, totally distributed with no
particles in the condensate. In spite of this failure, we conclude, however, that
our first objective of the work has been realized, namely our
T-matrix approach which has been applied for the first time to the
in-medium B-F case in \cite{Storo} seems to work reasonably well. 
In addition, for $t_b \neq t_f$, our formalism is still valid.
We compare in Table~\ref{table_one} the ground-state energies obtained 
by RPA, HF, and QMC for $t_f = 4 t_b$, and see that the results 
are still of the same quality as for the case with equal tunneling 
amplitude (see Fig.~\ref{GSE70}). Similar conclusions hold for the 
occupation numbers and for a comparison with exact 
 diagonalization for  a system size $L=6$.
\begin{table}[h!]
\centering
\begin{tabular}{|c|c|c|c|}
\hline
($U_{bf}$;$U_{bb}$) & $E_{RPA}$ & $E_{HF}$ & $E_{QMC}$\\
\hline
(-40;6) & -24.30 & -12.85 & -20.78\\
(-50;10)& -29.71 & -14.88 & -25.54\\
(-60;12)& -35.69 & -17.15 & -30.33\\
\hline
\end{tabular}
\caption{\label{table_one} Comparison of the ground state energies per site 
obtained by RPA, HF and QMC for a system of size $L=30$ and tunneling 
amplitudes $t_f = 4t_b$. The HF and RPA relatives errors are roughly constant.}
\end{table}
We also should note that we eventually would like to
apply our theory to the 3D case where mean field and RPA theories
usually perform much better, and where no QMC data are available.

\begin{figure*}[tph]
\epsfig{file=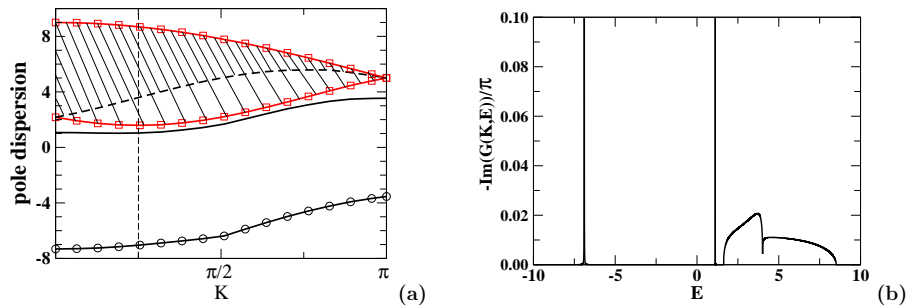,width=0.70\textwidth}
\caption{\label{spectral}(Color online) (a) The solid line with circles
represents the free fermion-like branch of the spectral function,
the solid line the second B-F branch, the solid lines with squares
the limits of the continuum and the dashed line a second plateau
in the continuum, (b) The spectral representation corresponding to
($\ref{GKE}$) for $\mathbf{K}=\pi/4$, see vertical broken line of
(a) ($U_{bf}=-10$, $k_F=\pi/4$).}
\end{figure*}

Another objective of this work is to confirm a surprising finding
discussed in \cite{Storo}, the appearance of a second stable B-F
branch at arbitrary small B-F (attractive) interaction. It exists
only because of the presence of a sharp Fermi surface and is thus
analogous to the formation of Cooper pairs in pure Fermi systems.
For this investigation, we pass to the continuum limit in which
the B-F propagator in HF approximation is given analytically by
\vspace{-0.1cm} \ba
G_K^0(E) = %\int_{-\pi}^{\pi} \frac{dk}{2\pi L} \frac{\Theta (\varepsilon ^f(k) - \varepsilon _f) + 2\pi n_0 \delta (K-k)}{E-\tilde{\varepsilon} ^f(k)-
%\tilde{\varepsilon} ^b(K-k)+i\eta}\\\nonumber
\frac{x}{\pi LP^0} \frac{1}{\sqrt {x^2-1}} \Bigl{(} \arctan{ \Bigl{(} \frac{(x+1)\cot (\frac{K}{4}+\frac{k_f}{2})}{\sqrt{x^2-1}} \Bigr{)}}
\hspace{1cm} \nonumber\\ (10)\hspace{1cm}\nonumber\\
-\arctan{ \Bigl{(} \frac{(x+1)\cot
(\frac{k_f}{2}-\frac{K}{4})}{\sqrt{x^2-1}} \Bigr{)}}\Bigr{)}
+\frac{n_0/L}{P^1+2+2\cos{(K)}} ~,\hspace{1cm} \nonumber \ea
\noindent with \ban
x&=&\frac{P^0}{4\cos{(\frac{K}{2})}} ~,\\
P^0&=&E+\frac{2k_f}{\pi}(U^{bb}+U^{bf})+i\eta ~,\\
P^1&=&E+\frac{k_f}{\pi}(U^{bb}+U^{bf})+i\eta ~. \ean

In Fig.\ref{spectral} we show the spectral function, i.e.
$-Im(G(K,E))/\pi$, for a typical set of parameters for
$N_B=N_F=L/4$, away from half filling. Besides the low lying peak
which corresponds to the free fermion dispersion in the
$U_{bf}\rightarrow 0$ limit, we see the striking feature that a
second stable branch develops right below the continuum. This
stable second branch exists for arbitrarily small values of $U_{bf}$. It
is the same phenomenon as was seen in our earlier work
\cite{Storo}. The existence of a Fermi surface entails a
logarithmic divergence of $Re(G^0)$ at $E /2=\varepsilon_F$ and
then there is always a sharp state below the continuum solution of
$1-g_{bf} G^0(\mathbf{K},E)=0$. However contrary to what happens
in the homogeneous continuum case \cite{Storo}, this second branch
does not interact strongly with the lowest free fermion-like
branch. In \cite{Storo} there was a level crossing of the two
branches which does not occur here. Also the upper branch has
rather little spectral weight compared with the lower one. It
would be interesting to see whether this second branch can be
found experimentally. In 1D, however, we should note that this
slightly detached second branch is certainly an artefact of the
T-matrix approximation because the correlated fermion occupation
numbers do not show any discontinuity (see fig. 3).   However in
2D or 3D, we think that this second branch should exist. Whether
it survives in a trap geometry \cite{pollet2} remains to be seen.

In conclusion, we have investigated in this work boson-fermion
pairing in a Bose-Fermi mixture on a 1D optical lattice. The
in-medium Boson-Fermion scattering problem was solved in T-matrix
approximation. As in a previous investigation, in the
thermodynamical limit, two stable branches were found. One
corresponds to the elastic scattering of the fermions off the Bose
condensate and the other is created from scattering of bosons out
of the condensate with fermions above the Fermi sea. The latter
comes because of the presence of a sharp Fermi surface and
therefore has the same origin as the Cooper pole in a pure two
component Fermi system. While in 1D systems, the second branch is
unphysical and just an artefact of the method, we think that in
3D such a branch should be real. We checked the validity of our
approach versus exact results available for a finite number of
sites. For most quantities we found good qualitative and
semi-quantitative agreement. This is satisfying because, as
already mentioned, RPA generally works better in higher
dimensions. More elaborate studies of B-F pairing are under way.\\
\\
Ongoing collaboration and discussions with T. Suzuki and C. Martin on B-F correlations 
are appreciated. We gratefully acknowledge contributions by J. Dukelsky during the
early stages of this work, the financial support of the Swiss
National Science Foundation and the partial support of the US
National Science Foundation under grant \# 0553127. Part of the
calculations discussed in this work were carried out on the
Hreidar cluster at ETH Zurich.

\end{document}